\documentclass[conference]{IEEEtran}

\usepackage[linesnumbered,lined, algoruled]{algorithm2e}
\usepackage{amssymb,bm}
\usepackage{amsmath}
\usepackage[acronym,shortcuts]{glossaries}
\usepackage{algorithmic,float}
\usepackage{balance}
\usepackage{color}
\usepackage{dsfont}
\usepackage{soul}
\usepackage{graphicx}
\usepackage{comment}
\usepackage{mathtools}
\usepackage{placeins}

\graphicspath{{./figures/}}

\makeatletter
\newsavebox\myboxA
\newsavebox\myboxB
\newlength\mylenA
\newcommand*\xoverline[2][0.75]{%
	\sbox{\myboxA}{$\m@th#2$}%
	\setbox\myboxB\null
	\ht\myboxB=\ht\myboxA%
	\dp\myboxB=\dp\myboxA%
	\wd\myboxB=#1\wd\myboxA
	\sbox\myboxB{$\m@th\overline{\copy\myboxB}$}
	\setlength\mylenA{\the\wd\myboxA}
	\addtolength\mylenA{-\the\wd\myboxB}%
	\ifdim\wd\myboxB<\wd\myboxA%
	\rlap{\hskip 0.5\mylenA\usebox\myboxB}{\usebox\myboxA}%
	\else
	\hskip -0.5\mylenA\rlap{\usebox\myboxA}{\hskip 0.5\mylenA\usebox\myboxB}%
	\fi}
\makeatother

\makeglossaries
\newacronym{3gpp}{3GPP}{3rd Generation Partnership Project}
\newacronym{5g}{5G}{fifth generation}
\newacronym{6g}{6G}{sixth generation}
\newacronym{awgn}{AWGN}{additive white Gaussian noise}
\newacronym{bler}{BLER}{block error rate}
\newacronym{bs}{BS}{base station}
\newacronym{ecf}{ECF}{empirical characteristic function}
\newacronym{cdf}{CDF}{cumulative distribution function}
\newacronym{fa}{FA}{false alarm}
\newacronym{glrt}{GLRT}{generalized likelihood ratio test}
\newacronym{kpi}{KPI}{key performance indicator}
\newacronym{los}{LOS}{line of sight}
\newacronym{lte}{LTE}{long term evolution}
\newacronym{mbb}{MBB}{mobile broadband}
\newacronym{md}{MD}{missed detection}
\newacronym{mimo}{MIMO}{multiple-input multiple-output}
\newacronym{mle}{MLE}{maximum likelihood estimate}
\newacronym{mse}{MSE}{mean squared error}
\newacronym{np}{NP}{Neyman-Pearson}
\newacronym{nr}{NR}{new radio}
\newacronym{ofdm}{OFDM}{orthogonal frequency division multiplexing}
\newacronym{pdcp}{PDCP}{packet data convergence protocol}
\newacronym{pdf}{PDF}{probability density function}
\newacronym{pmf}{PMF}{probability mass function}
\newacronym{prb}{PRB}{physical resource block}
\newacronym{re}{RE}{resource element}
\newacronym{rf}{RF}{radio frequency}
\newacronym{roc}{ROC}{receiver operating characteristic}
\newacronym{rv}{r.v.}{random variable}
\newacronym{se}{SE}{spectral efficiency}
\newacronym{sinr}{SINR}{signal to interference plus noise ratio}
\newacronym{siso}{SISO}{single-input single-output}
\newacronym{snr}{SNR}{signal to noise ratio}
\newacronym{ue}{UE}{user equipment}
\newacronym{ul}{UL}{uplink}
\newacronym{urllc}{URLLC}{ultra-reliable low-latency communications}

\begin{document}

\title{Jamming Detection with Subcarrier Blanking for 5G and Beyond in Industry 4.0 Scenarios}

\author{Leonardo Chiarello$^{*\mathsection}$, Paolo Baracca$^\mathsection$, Karthik Upadhya$^\dagger$, Saeed R. Khosravirad$^\ddagger$, and Thorsten Wild$^\mathsection$\\
$^*$Department of Information Engineering, University of Padova, Italy\\
$^\mathsection$Nokia Bell Labs, Stuttgart, Germany\\
$^\dagger$Nokia Bell Labs, Espoo, Finland\\
$^\ddagger$Nokia Bell Labs, Murray Hill, USA\\}

\maketitle

\sloppy
\begin{abstract}
Security attacks at the physical layer, in the form of radio jamming for denial of service, are an increasing threat in the Industry 4.0 scenarios. In this paper, we consider the problem of jamming detection in \acs{5g}-and-beyond communication systems and propose a defense mechanism based on pseudo-random blanking of subcarriers with \ac{ofdm}. We then design a detector by applying the \ac{glrt} on those subcarriers. We finally evaluate the performance of the proposed technique against a smart jammer, which is pursuing one of the following objectives: maximize stealthiness, minimize \ac{se} with \ac{mbb} type of traffic, and maximize \ac{bler} with \ac{urllc}. Numerical results 
show that a smart jammer a) needs to compromise between \ac{md} probability and \ac{se} reduction with \ac{mbb} and b) can achieve low detectability and high system performance degradation with \ac{urllc} only if it has sufficiently high power.

\end{abstract}


\glsresetall

\section{Introduction}\label{sec:intro}

Security has been one of the main pillars driving the \ac{3gpp} design of the \ac{5g} of mobile communication systems. In fact, several security functionalities are available in \ac{5g} at the \ac{pdcp} and above layers to guarantee authentication, privacy and data integrity \cite{dahlman_2018}. On the other hand, denial of service attacks in the form of radio jamming have been recently recognized as a major threat for the \ac{5g} deployment in Industry 4.0 scenarios, in particular with \ac{urllc}
. As an example, while we can assume that no malicious device can be activated inside a factory, it might happen that a jammer stationed outside the plant blocks the transmission of some legitimate devices inside that plant. Such attack can cause large economic losses to the factory by interrupting the production. Moreover, besides these \ac{5g} Industry 4.0 use cases, jamming detection and mitigation has been recognized as an extremely relevant topic also for \ac{6g} technologies \cite{viswanathan_2020}.

Jamming attacks have been known as a threat for communication and localization systems for many years, and jammers have been extensively used in the military context to degrade the effectiveness of enemy radars. Mainly for that reason, it is very simple and inexpensive nowadays to obtain a jammer that is capable of emitting high jamming power 
\cite{jammerstore_2021}. Furthermore, aside from simple devices that can generate narrow- or wide-bands \ac{rf} interference \cite{xu_2006}, there exist much smarter but still easily available jamming devices too \cite{wilhelm_2011}. This last type, a.k.a. as {\it reactive} jammers, are inactive while no legitimate transmission is happening, and then starts generating interference as soon as they sense some transmission on the channel, making them very difficult to be detected
. Differently from other security aspects like authentication and privacy that can and are well managed at \ac{pdcp} and above layers in \ac{5g}, jamming, as a form of malicious interference, can be handled at the physical layer. In fact, physical layer security mechanisms are expected to play an important role in \ac{6g} \cite{chorti_2021}.

A fundamental difference exists between a legitimate interfering device and a jammer. A legitimate device creates interference while respecting the rules of the standard regulating communications in that band and several well-known techniques exist to deal with that type of interference. A jammer is a malicious device that intentionally attacks the system, also violating the regulatory and standardization rules, and its activity can be extremely dangerous 
\cite{orakcal_2012}. For that reason, a jamming-resilient communication system needs to perform two tasks: a) detection, to understand that some network performance degradation happens because of a malicious jamming attack and not because of fading or some legitimate cellular interference, and b) mitigation, with the implementation of focused techniques to limit the impact of the attacker.

Here, we consider the problem of jamming detection in \ac{5g}-and-beyond communication systems, with particular focus on \ac{urllc}. Some work has recently been done in this framework. In \cite{do_2018}, a detection technique has been proposed for massive \ac{mimo} \acp{bs} exploiting pseudo-random hopping of the scheduled \acp{ue} among the pilot sequences and allowing the \ac{bs} to design a jamming-resilient combiner. A more specific analysis for \ac{urllc} has been done by \cite{swamy_2019}, where an advanced feedback is proposed and relays implementing promiscuous listening are used to detect rare events like a jammer. Along this direction, \cite{zhang_2018} proposes to deploy a guard node that generates and transmits a signal known at the legitimate receiver: the received signal is then post-processed to determine if a jamming attack occurred.

In this work we consider a system using \ac{ofdm} and propose a novel method based on pseudo-random blanking of subcarriers to detect jamming attacks. Differently from \cite{swamy_2019, zhang_2018}, our proposal does not require the deployment of additional nodes and, differently from \cite{do_2018}, it applies also to the case of \acp{bs} with limited number of antennas, which occurs in particular with \ac{5g} deployments for Industry 4.0 indoor scenarios
. Moreover, our proposal can be seamlessly embedded in the air interface design of \acs{ofdm}-based technologies, including the \ac{lte}, \ac{5g} \ac{nr} and the anticipated \ac{6g}.
More specifically, we design a detector exploiting the \ac{glrt} and aiming to detect a smart jammer. Namely, we study a jammer with one of the following objectives: a) maximize the \ac{md} probability to remain stealthy, b) minimize the \ac{se}, assuming a \ac{mbb} type of traffic, and c) maximize the \ac{bler}, considering a \ac{urllc} type of traffic. The subcarrier blanking represents a loss in terms of system performance as some resources are not used for communication. Numerical results are provided to show the benefits of the proposed approach and the trade-off between the capability of detecting a jammer and the system performance
.

\emph{Notation}. $(\cdot)^\textnormal{T}$ denotes the transpose. $\lVert \textbf{X} \rVert$ is the Euclidean norm of $\textbf{X}$. $|\mathcal{X}|$ denotes the cardinality of the set $\mathcal{X}$. $\xoverline{\mathcal{X}}$ denotes the complement of the set $\mathcal{X}$. $\textnormal{diag}(x_1,\dots,x_N)$ denotes the diagonal matrix where $x_1,\dots,x_N$ are the diagonal elements. $C_{n,k} = \binom{n}{k} = \frac{n!}{k!(n-k)!}$ denotes the number of $k$-combinations from a given set of $n$ elements. $\mathcal{CN}(\mu,\sigma^2)$ denotes the complex Gaussian distribution with mean $\mu$ and variance $\sigma^2$. $Q(\cdot)$ denotes the Gaussian Q-function. $P[A]$ denotes the probability of event $A$. $\mathbb{E}[X]$ denotes the expectation of \ac{rv} $X$.
\section{System Model}\label{sec:system_model}

We consider a single-cell single-user uplink scenario with a \ac{ue} transmitting toward a \ac{bs}; both \ac{ue} and \ac{bs} are single-antenna. Moreover, we assume a jammer that tries to disrupt the ongoing communication by sending a malicious signal to the \ac{bs}
. The considered system uses an \ac{ofdm} modulation where the available radio resources can be thought as in a resource grid composed of \acp{re}, where each \ac{re} occupies one subcarrier in frequency and one OFDM symbol in time, for a total of $S$ subcarriers per OFDM symbol (see also Fig. \ref{fig:ofdm_grid}).


We now define the signal received by the \ac{bs} on subcarrier $s \in \{ 1, \dots, S \}$ at symbol $n \in\mathbb{N}$ as
\begin{equation}
	\label{eq:received_signal}
	r_s(n) = u_s(n) \cdot h_s(n) + j_s(n) \cdot g_s(n) + w_s(n) \, ,
\end{equation}
where $u_s(n)$ is the signal from the \ac{ue}, $h_s(n)$ is the \ac{ue} channel, $j_s(n)$ is the signal from the jammer, and $g_s(n)$ is the jammer channel. Moreover, $w_s(n) \sim \mathcal{CN}(0,\sigma_w^2)$ is the complex Gaussian noise with $\sigma_w^2$ as statistical power. In particular, in this paper we consider two types of channel:
\begin{itemize}
	\item \ac{awgn} channel with $h_s(n) = H$ and $g_s(n) = G$, where $H$ and $G$ are constant for each subcarrier and OFDM symbol;
	\item Rayleigh channel with $h_s(n) \sim \mathcal{CN}(0,\sigma_h^2)$ and $g_s(n) \sim \mathcal{CN}(0,\sigma_g^2)$.
\end{itemize}
We can now write the corresponding \ac{sinr} on subcarrier $s \in \{ 1, \dots, S \}$ at symbol $n \in \mathbb{N}$ as
\begin{equation}
	\textnormal{SINR}_s(n) = \frac{P_{\textnormal{UE},s}(n) \lVert h_s(n) \rVert^2 }{\sigma_w^2 + P_{\textnormal{J},s}(n) \lVert g_s(n) \rVert^2} \, ,
\end{equation}
where $P_{\textnormal{UE},s}(n)$ is the \ac{ue} power and $P_{\textnormal{J},s}(n)$ is the jammer power. We denote the respective total power per symbol as $P_{\textnormal{UE}}(n) = \sum_{s=1}^{S} P_{\textnormal{UE},s}(n)$ and $P_{\textnormal{J}}(n) = \sum_{s=1}^{S} P_{\textnormal{J},s}(n)$.

Regarding the \acp{kpi} that will be used to evaluate the damage caused by the jammer to the legitimate system, in this work we focus on \ac{se} and \ac{bler}. \ac{se} is most relevant when considering \ac{mbb} type of traffic and we define it as
\begin{equation}
	\textnormal{SE} = \mathbb{E} \left[ \log_2 \left( 1 + \textnormal{SINR}_s(n) \right) \right] \, .
\end{equation}
For what concerns the \ac{bler}, it comes in handy for evaluating the system performance in a \ac{urllc} type of traffic. In fact, in the \ac{urllc} case we assume that we have small packets sent by the \ac{ue}, each packet scheduled on a set of \acp{re} allocated within a limited number of \ac{ofdm} symbols: because of the tight latency constraint of \ac{urllc}, we do not assume retransmission capabilities. For our analysis, we define a certain $\textnormal{SINR}_{\textnormal{pkt}}$ as the equivalent \ac{sinr} experienced by a packet (which can be computed by using different link-to-system mapping criteria \cite{brueninghaus_2005}), and using this we 
compute the \ac{bler} 
from the normal approximation of the finite blocklength capacity:
\begin{equation}
	\label{eq:bler}
	\textnormal{BLER}_{\textnormal{pkt}} = Q \left( \left[ \log_2 \left( 1 + \textnormal{SINR}_{\textnormal{pkt}} \right) - \rho \right] \sqrt{\frac{C}{V}} \right)
\end{equation}
where
$V$
is the channel dispersion, $\rho$ is the packet spectral efficiency, and $C$ is the coded packet size \cite[Eq. (5)]{durisi_2016}.
\section{Defense Strategy}\label{sec:defense_strategy}

\begin{figure}
	\centering
	\includegraphics[width=0.4\textwidth]{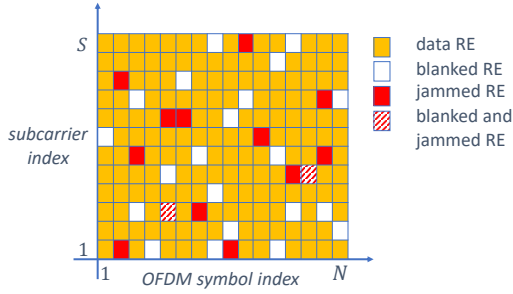}
	\caption{Example of the \ac{ofdm} resource grid with blanked and jammed \acp{re}.}
	\label{fig:ofdm_grid}
\end{figure}

In this work, we propose to blank some \acp{re} in each \ac{ofdm} symbol in a pseudo-random manner, such that the attacker cannot predict in advance which resources will be used for transmission and which will be blanked. 
With our proposal, we have then two types of \acp{re}: data \acp{re} and blanked \acp{re}. The former is used for data transmission, while the latter is, indeed, left blanked and will be used for jamming detection. In particular, at each OFDM symbol $n$, the \ac{ue} blanks a set $\mathcal{M}(n) = \{ m_1(n), \dots, m_M(n) \}$ (with cardinality $M = |\mathcal{M}(n)|$) of \acp{re}, where $m_1(n), \dots, m_M(n)$ are chosen in a pseudo-random manner from the set $\{ 1, \dots, S \}$; the remaining \acp{re} are used for data transmission. Since no data is transmitted on the blanked \acp{re}, we define the \ac{ue} signal introduced in (\ref{eq:received_signal}) as
\begin{equation}
	u_s(n) = 
	\begin{cases}
		d_s(n) & s \in \xoverline{\mathcal{M}(n)} \\
		0  & s \in \mathcal{M}(n) \\
	\end{cases} \, ,
	\qquad n\in \mathbb{N} \, ,
\end{equation}
where $d_s(n)$ is the data sample sent by the \ac{ue}. At the same time, the attacker transmits on a set $\mathcal{L}(n) = \{ \ell_1(n), \dots, \ell_L(n) \}$ (with cardinality $L = |\mathcal{L}(n)|$) of \acp{re}, where $\ell_1(n), \dots, \ell_L(n)$ are chosen from the set $\{ 1, \dots, S \}$ according to the jammer strategy. Note that while here we consider mainly for sake of notation $M$ and $L$ constant, in practice they can also be discrete random variables. Fig. \ref{fig:ofdm_grid} shows an example of the resource grid in such situation.

The defense strategy takes advantage of the blanked \acp{re} to detect the presence of jamming 
by means of statistical hypothesis testing \cite{kay_1993}. The two hypotheses for the sequence of blanked \acp{re} are as follows:
\begin{itemize}
	\item There is no jamming and we have just thermal noise (null hypothesis $\mathcal{H}_0$);
	\item There is jamming (alternative hypothesis $\mathcal{H}_1$).
\end{itemize}
Although we consider here a single-cell scenario, this method can be extended to the case with multiple BSs that share the same sequence of REs to be blanked. That is particularly beneficial for instance in Industry 4.0 scenarios in the several countries where specific bands are now being allocated to industry players for deploying their own 5G-and-beyond network without interference from neighbouring networks \cite{bundesnetzagentur_2019}.

By denoting with $N$ the number of \ac{ofdm} symbols that will be used at the \ac{bs} for jamming detection, the above hypotheses translate in the following hypothesis test:
\begin{equation}
	\begin{cases}
		\mathcal{H}_0: \mathbf{r} = \mathbf{w}\\
		\mathcal{H}_1: \mathbf{r} = \mathbf{w} + \mathbf{G} \mathbf{j}
	\end{cases} \, ,
\end{equation}
where $\mathbf{r} = \left[ r_{m_1(1)} \cdots r_{m_M(1)} \cdots r_{m_1(N)} \cdots r_{m_M(N)} \right]^\textnormal{T}$, $\mathbf{w} = \left[ w_{m_1(1)} \cdots w_{m_M(1)} \cdots w_{m_1(N)} \cdots w_{m_M(N)} \right]^\textnormal{T}$, and $\mathbf{j} = \left[ j_{m_1(1)} \cdots j_{m_M(1)} \cdots j_{m_1(N)} \cdots j_{m_M(N)} \right]^\textnormal{T}$. These are vectors containing the samples of all the blanked \acp{re} for, respectively, received, noise, and jamming signal. Moreover, $\mathbf{G} = \textnormal{diag}(g_{m_1(1)}, \dots,g_{m_M(1)}, \dots, g_{m_1(N)}, \dots, g_{m_M(N)})$ is the jammer channel matrix.

After defining the received jamming signal as $\tilde{\mathbf{j}} = \mathbf{G} \mathbf{j}$, we propose here a detector in the most general case where no assumption can be made about $\tilde{\mathbf{j}}$. Then, in Section \ref{sec:gaussian_jammer}, we use the derived detector for computing the \ac{md} probability in closed form in the case that $\tilde{\mathbf{j}}$ has zero-mean complex Gaussian distribution.

By assuming that the statistics of the jamming signal received by the \ac{bs} are unknown, $\mathcal{H}_1$ becomes a composite alternative hypothesis and we need to resort to the \ac{glrt} \cite{kay_1993}. The \ac{glrt} decides for $\mathcal{H}_1$ if
\begin{equation}
	\label{eq:decision_rule_1}
	\Lambda(\mathbf{r}) = \frac{p(\mathbf{r} | \hat{\mathbf{j}}; \mathcal{H}_1)}{p(\mathbf{r}; \mathcal{H}_0)} > \lambda \, ,
\end{equation}
where $p(\mathbf{r}; \mathcal{H}_0)$ is the \ac{pdf} of $\mathbf{r}$  under $\mathcal{H}_0$ and $p(\mathbf{r} | \hat{\mathbf{j}}; \mathcal{H}_1)$ is the \ac{pdf} of $\mathbf{r}$ conditioned on $\tilde{\mathbf{j}} = \hat{\mathbf{j}}$ and under $\mathcal{H}_1$. Moreover, $\hat{\mathbf{j}}$ is the \ac{mle} of $\tilde{\mathbf{j}}$ assuming $\mathcal{H}_1$ is true, $\Lambda(\mathbf{r})$ is the test statistic, and $\lambda$ is the threshold. In particular, $\lambda$ is found from
\begin{equation}
	\label{eq:pfa}
	P_{\textnormal{FA}}= P \left[ \Lambda(\mathbf{r}) > \lambda; \mathcal{H}_0 \right] = \int_{\left\{ \mathbf{r}: \Lambda(\mathbf{r}) > \lambda \right\}} p(\mathbf{r}; \mathcal{H}_0) d\mathbf{r} \, ,
\end{equation}
where $P_{\textnormal{FA}}$ is the \ac{fa} probability, i.e., the probability of declaring jamming even if it is not present.

We now derive the threshold $\lambda$ by fixing the value of $P_{\textnormal{FA}}$. First of all, we need to compute the test statistic formula, and we start by deriving the \ac{pdf} of $\mathbf{r}$ under $\mathcal{H}_0$ and $\mathcal{H}_1$. When $\mathcal{H}_0$ is true, the signal received at the \ac{bs} is $\mathbf{r} = \mathbf{w}$, therefore, we have $\mathbf{r} \sim \mathcal{CN}(\mathbf{0}, \sigma_w^2 \mathbf{I})$. When $\mathcal{H}_1$ is true, the received signal becomes $\mathbf{r} = \mathbf{w} + \tilde{\mathbf{j}}$, where $\tilde{\mathbf{j}}$ is an unknown vector, with resulting distribution $\mathbf{r} \sim \mathcal{CN}(\tilde{\mathbf{j}}, \sigma_w^2 \mathbf{I})$. In order to derive the corresponding \ac{pdf}, we need to compute the \ac{mle} of $\tilde{\mathbf{j}}$ by maximizing $p(\mathbf{r} | \tilde{\mathbf{j}}; \mathcal{H}_1)$ through the following optimization problem:
\begin{equation}
	\hat{\mathbf{j}} = \underset{\tilde{\mathbf{j}} \in \mathbb{C}^{M N \times 1}}{\arg\max} \, p(\mathbf{r} | \tilde{\mathbf{j}}; \mathcal{H}_1) \, .
\end{equation}
Under our assumptions, the solution is just $\hat{\mathbf{j}} = \mathbf{r}$, thus leading to
\begin{equation}
	\label{eq:pdf_H1}
	p(\mathbf{r} | \hat{\mathbf{j}}; \mathcal{H}_1) = p(\mathbf{r} | \mathbf{r}; \mathcal{H}_1) = \frac{1}{\left( \pi \sigma_w^2 \right)^{M N}} \, .
\end{equation}
By applying (\ref{eq:pdf_H1}) to (\ref{eq:decision_rule_1}) 
and after some computations, the resulting decision rule can be rewritten as
\begin{equation}
	\label{eq:decision_rule_2}
	\Lambda'(\mathbf{r}) = \frac{\lVert \mathbf{r} \rVert^2}{M N} > \lambda' \, ,
\end{equation}
with $\Lambda'(\mathbf{r}) \triangleq \frac{\sigma_w^2 \ln{\Lambda(\mathbf{r})}}{M N}$ being the new test statistic and $\lambda' \triangleq \frac{\sigma_w^2 \ln{\lambda}}{M N}$ the new threshold. This is, basically, an energy detector, which is quite an intuitive result: when we have no knowledge about the jamming signal, we can just compute the energy of the received signal on the blanked \acp{re} and compare it against a thereshold.

The test statistic distribution under $\mathcal{H}_0$ is then
\begin{equation}
	\label{eq:test_statistic_distribution}
	\Lambda'(\mathbf{r}; \mathcal{H}_0) = \frac{\lVert \mathbf{w} \rVert^2}{M N} \sim \textnormal{Gamma} \left( M N, \frac{\sigma_w^2}{M N} \right) \, ,
\end{equation}
where $\textnormal{Gamma}(k, \theta)$ is the gamma distribution with shape paramenter $k$ and scale parameter $\theta$. From (\ref{eq:pfa}), the resulting \ac{fa} probability turns out to be
\begin{equation}
		P_{\textnormal{FA}}= P \left[ \Lambda'(\mathbf{r}; \mathcal{H}_0) > \lambda' \right] = 1 - F_{	\Lambda'(\mathbf{r}; \mathcal{H}_0)}(\lambda') \, ,
\end{equation}
where $F_X(x)$ denotes the \ac{cdf} of the \ac{rv} $X$ computed in $x$. This leads to
\begin{equation}
	\label{eq:threshold}
	\lambda' = F^{-1}_{\Lambda'(\mathbf{r}; \mathcal{H}_0)} \left( 1 - P_{\textnormal{FA}} \right) \, ,
\end{equation}
which can be now used to evaluate the detection performance of this defense strategy. This is done by means of the \ac{md} probability, defined as the probability of accepting $\mathcal{H}_0$ when jamming is present. In this model, the test statistic distribution under $\mathcal{H}_1$ can be written as
\begin{equation}
	\label{eq:test_statistic_H1}
	\Lambda'(\mathbf{r}; \mathcal{H}_1) = \frac{\lVert \mathbf{w} + \tilde{\mathbf{j}} \rVert^2}{M N} \sim	\frac{\sigma_w^2}{2 M N} \chi^2 \left(2 M N, \lVert \tilde{\mathbf{j}} \rVert^2 \right) \, ,
\end{equation}
where $\chi^2(\nu, \delta)$ is the non-central $\chi^2$ distribution with $\nu$ degrees of freedom and non-centrality parameter $\delta$.
Eventually, we can write the \ac{md} probability as
\begin{equation}
	P_{\textnormal{MD}} = P \left[ \Lambda'(\mathbf{r}; \mathcal{H}_1) < \lambda' \right]  \nonumber = 1 - F_{ \Lambda'(\mathbf{r}; \mathcal{H}_1)}\left(\lambda' \frac{2 M N}{\sigma_w^2}\right) \, .
\end{equation}

\section{MD Probability with Gaussian Distributed Received Jamming Signal}\label{sec:gaussian_jammer}

We now evaluate the effectiveness of the proposed defense strategy against a jamming signal with distribution $\tilde{\mathbf{j}} \sim  \mathcal{CN} \left( \mathbf{0}, \mathbf{D} \right)$, where $\mathbf{D} = \textnormal{diag}(d_{m_1(1)}, \dots,d_{m_M(1)}, \dots, d_{m_1(N)}, \dots, d_{m_M(N)})$ is the covariance matrix with diagonal elements defined as
\begin{equation}
	d_{m_i(n)} =
	\begin{cases}
		P_{\textnormal{J}}/L, & m_i(n) \in \mathcal{L}(n) \\
		0 & m_i(n) \in \xoverline{\mathcal{L}(n)}
	\end{cases} \, ,
\end{equation}
with $P_{\textnormal{J}}$ the jamming power per symbol, $i = 1,\dots,M$, and $n = 1,\dots,N$. In order to evaluate the performance of this type of attack against the defense mechanism, we apply the decision rule defined in (\ref{eq:decision_rule_2}) and we compute in closed form the corresponding \ac{md} probability. This result, besides for the analysis purpose, will be useful in Section \ref{sec:jamming_strategies} when deriving the best jammer strategy for minimizing its detectability.

For this computation, we need to take into account that the \ac{md} probability at a given time depends on the number of jammed \acp{re} that falls into the blanked ones. Therefore, first of all, we define the set $\mathcal{E} = \left[ \mathcal{M}(1) \cap \mathcal{L}(1) \right] \cup \dots \cup \left[ \mathcal{M}(N) \cap \mathcal{L}(N) \right] = \{ e_1, \dots, e_E \}$ (with cardinality $E = |\mathcal{E}|$) of overlapping blanked and jammed \acp{re}. Then, by denoting with $\widetilde{\Lambda}'(\mathbf{r}, E; \mathcal{H}_1)$ the test statistic under $\mathcal{H}_1$ in (\ref{eq:test_statistic_H1}) with the new assumption of Gaussian jammer,  the \ac{md} probability can be computed as
\begin{equation}
	\label{eq:pmd_tpt}
	\widetilde{P}_{\textnormal{MD}} = \sum_{e = E_{\textnormal{min}}}^{E_{\textnormal{max}}} P \left[ \widetilde{\Lambda}'(\mathbf{r}, E; \mathcal{H}_1) < \lambda' | E = e \right] P \left[ E = e \right] \, ,
\end{equation}
where the law of total probability has been applied. Moreover,
\begin{align}
	E_{\textnormal{max}} &= \min (MN,LN) \, , \\
	E_{\textnormal{min}} &= 
	\begin{cases}
		0 & \text{if } M + L < S \\ 
		(M + L - S)N & \text{if } M + L \geq S
	\end{cases} \, ,
\end{align}
are the minimum and maximum number of overlapping \acp{re}, given $M$, $L$, and $S$. We now need to derive the two factors in (\ref{eq:pmd_tpt}), i.e., the \ac{cdf} of the test statistic under $\mathcal{H}_1$ and the \ac{pmf} of $E$.

Starting from the former, the test statistic distribution under $\mathcal{H}_1$, after some computations, results in
\begin{equation}
		\widetilde{\Lambda}'(\mathbf{r}, E; \mathcal{H}_1) \sim E \cdot \mathcal{E} \left( \frac{M N}{\sigma_w^2 + \sigma_j^2} \right) + (MN - E) \cdot \mathcal{E} \left( \frac{M N}{\sigma_w^2} \right) \, ,
\end{equation}
where $\mathcal{E}(1/\beta)$ is the exponential distribution with rate parameter $1/\beta$. To derive its \ac{cdf}, we take advantage of a result in \cite[Eq. (9)]{chaitanya_2013} on the \ac{cdf} of the sum of independent exponential \acp{rv}:
\begin{equation}
		P \left[ \widetilde{\Lambda}'(\mathbf{r}, E; \mathcal{H}_1) < \lambda' \right] = 1 - \sum_{i=1}^{2} \sum_{j=1}^{\alpha_i} \sum_{k=0}^{j-1} \frac{\chi_{i,j} (\lambda')^k}{k! \beta^k_{\langle i \rangle}}  \textnormal{e}^{-\frac{\lambda'}{\beta_{\langle i \rangle}}} \, ,
\end{equation}
where $\alpha_1 = E$, $\alpha_2 = MN - E$, $\beta_{\langle 1 \rangle} = \frac{\sigma_w^2 + \sigma_j^2}{M N}$, $\beta_{\langle 2 \rangle} = \frac{\sigma_w^2}{M N}$, and
\begin{equation*}
		\chi_{i,j} = \left( - \frac{1}{\beta_{\langle i \rangle}} \right)^{\omega_{i,j}} \cdot C_{\alpha_b + \omega_{i,j} - 1,\omega_{i,j}} \cdot \frac{\beta_{\langle b \rangle}^{\omega_{i,j}}}{\left( 1 - \frac{\beta_{\langle b \rangle}}{\beta_{\langle i \rangle}} \right)^{\alpha_b + \omega_{i,j}}} \, ,
\end{equation*}
with $b \neq i$ and $\omega_{i,j} = \alpha_i - j$.

Finally, we observe that $E$ is a hypergeometric random variable, whose \ac{pmf} can be written as
\begin{equation}
	P \left[ E = e \right] = \frac{C_{(S-L)N,MN-e} C_{LN,e}}{C_{SN,MN}} \, ,
\end{equation}
where
\begin{itemize}
	\item $C_{SN,MN}$ is the number of ways to choose $MN$ total blanked subcarriers out of $SN$ total subcarriers;
	\item $C_{(S-L)N,MN-e}$ is the number of ways to choose $MN-e$ blanked subcarriers ($e$ overlapping subcarriers are fixed) out of $(S-L)N$ subcarriers ($LN$ jammed subcarriers are fixed);
	\item $C_{LN,e}$ is the number of ways to choose $e$ overlapping subcarriers out of $LN$ total jammed subcarriers.
\end{itemize}

\section{Jamming strategies}\label{sec:jamming_strategies}

A jammer has the two tasks of a) not being detected and b) minimize the system performance. Since expressing an optimization problem that considers both tasks at the same time is not trivial 
, we consider in this work the following three heuristic strategies in order to achieve the above objectives:
\begin{itemize}
	\item maximize the \ac{md} probability, to remain as much undetected as possible, but still transmitting at maximum power;
	\item minimize the \ac{se}, to reduce the performance with \ac{mbb} type of traffic;
	\item maximize the \ac{bler}, to disrupt a \ac{urllc} type of traffic.
\end{itemize}
Before going through all of them, we first make some assumptions on the jammer knowledge of the system. While it is fair that the jammer knows or can estimate many parameters (either because defined by the standard or just because it can listen to \ac{bs} and \ac{ue} transmission), some variables cannot be easily tracked by the attacker, like the instantaneous channel between the \ac{ue} and the \ac{bs}. Therefore, in general, we assume the jammer to know the format of the transmission
, the large scale fading, and the noise statistical power at the \ac{bs}. Moreover, regarding the defensive parameters, it is reasonable to assume that the jammer knows the number of blanked \acp{re} $M$, by estimating it (since we consider it fixed in time), and the number of symbols $N$ that the defense strategy uses for detection.

\subsection{\ac{md} Probability Maximization}\label{sec:jamming_strategies_A}

A jammer transmitting at maximum power and equally distributing it among the attacked subcarriers $L$ selected in a pseudo-random way, and that also wants to maximize the \ac{md} probability, needs to solve the following optimization problem
\begin{equation}
		L^* = \underset{1 \leq L \leq S}{\arg\max} \, \widetilde{P}_{\textnormal{MD}}(L) \, ,
\end{equation}
where $\widetilde{P}_{\textnormal{MD}}(L)$ has been computed in (\ref{eq:pmd_tpt}). Here we reasonably assume that $\lambda'$ is known by the jammer. The above optimization problem is not trivial to solve, mainly because the function to maximize is transcendental and $L$ is discrete and present in the bounds of the summation. However, since the objective function depends only on a single discrete variable, the attacker can solve it by performing an exhaustive search to find the optimal value.

\subsection{\ac{se} Minimization}\label{sec:jamming_strategies_B}

In this case, we assume the jammer to ignore the detectability problem and just try to minimize the system \ac{se}. Because of that, the jammer considers here an \ac{ofdm} system without any blanking and needs to solve the following optimization problem:
\begin{equation}
	\begin{split}
		\mathbf{P_{\textnormal{J}}}^* ={}& \underset{\mathbf{P_{\textnormal{J}}} \in \mathbb{R}^S}{\arg\min} \, \sum_{s=1}^{S} \log_2 \left( 1 + \frac{P_{\textnormal{UE},s} E_h}{\sigma_w^2 + P_{\textnormal{J},s} E_g } \right) \, , \\
		&\text{subject to } \sum_{s=1}^{S} P_{\textnormal{J},s} = P_{\textnormal{J}}
	\end{split}
\end{equation}
where $\mathbf{P_{\textnormal{J}}} = \left[ P_{\textnormal{J},1} \cdots P_{\textnormal{J},S} \right]$, and $E_h$ and $E_g$ are the average \ac{ue} and jammer channel energies, more specifically being $E_h = H^2$ and $E_g = G^2$ with \ac{awgn} channel, and $E_h = \sigma_h^2$ and $E_g = \sigma_g^2$ in the Rayleigh scenario. Moreover, we assume that $P_{\textnormal{UE},1} E_h = \cdots = P_{\textnormal{UE},S} E_h = \widehat{P}_{\textnormal{UE}}$.

It is straightforward to show, but not reported here for the sake of space, that the solution to the above problem can be obtained with the method of the Lagrange multipliers and is $P_{\textnormal{J},s} = \frac{P_{\textnormal{J}}}{S}$, i.e., if the jammer wants to minimize the \ac{se}, it must perform a wide-band attack.

\subsection{\ac{bler} Maximization}\label{sec:jamming_strategies_C}

In this attack we still assume the jammer to ignore the detectability issue and try to minimize the system performance, which on the other hand is measured with the \ac{bler} as \ac{kpi} (\ref{eq:bler}). A formulation of this problem is in general not straightforward at the jammer, as it would require the attacker to know how many and on which resources these packets are scheduled. So here, we consider a suboptimal approach where the jammer assumes $F$ packets scheduled on the $N$ \ac{ofdm} symbols, each packet allocated to $S/F$ neighbouring subcarriers with no spatial multiplexing, i.e., packets are scheduled next to each other in the frequency domain. Moreover, we still assume that the jammer transmits at full power with equal split among the attacked subcarriers, and if a packet is attacked all the subcarriers of that packet are jammed. Under these assumptions, the jammer just needs to determine how many packets to attack by solving the following optimization problem:
\begin{equation}
	\label{eq:bler_opt}
	L_{\textnormal{F}}^* = \underset{1 \leq L_{\textnormal{F}} \leq F}{\arg\max} \, L_{\textnormal{F}} \cdot \textnormal{BLER}_{\textnormal{pkt}} (P_J / L_F) \,,
\end{equation}
where $\textnormal{BLER}_{\textnormal{pkt}}(x)$ is the \ac{bler} computed from (\ref{eq:bler}) by considering in the \ac{sinr} computation average channel energy and $x$ as interference power.

Similarly to the \ac{md} probability maximization problem, the optimization problem (\ref{eq:bler_opt}) is not trivial to solve in closed form. However, as before, since the objective function depends only on a single discrete variable, the attacker can perform an exhaustive search to find the optimal value.

\section{Numerical Results}\label{sec:sim_results}

We consider an OFDM system with $S=300$ subcarriers, compliant to a \ac{5g} numerology with $60 \, \textnormal{kHz}$ as subcarrier spacing for a $20 \, \textnormal{MHz}$ bandwidth. The subcarriers are grouped into \acp{prb}, each consisting of $S_P = 12$ consecutive subcarriers, and transmission happens in slot of $14$ \ac{ofdm} symbols \cite{dahlman_2018}. Therefore, we have $P = S/S_P =2 5$ \acp{prb} per slot, and we assume the detection to be performed per slot, i.e., $N = 14$. As the \ac{prb} is the smallest time-frequency resource that can be scheduled to a device, we implement a \ac{5g} standard compliant defense with blanking performed per \ac{prb} and, as a consequence, introduce $M_P$ as the number of blanked \acp{prb} per slot. For simplicity, we also assume the jammer to perform attacks on a \ac{prb} basis and denote with $L_P$ the number of jammed \acp{prb} per slot. 

We introduce now some parameters to better define the considered simulation setup:
\begin{itemize}
	\item $\textnormal{SNR}_{\textnormal{UE}}(n) = P_{\textnormal{UE}}(n) / (S \cdot \sigma_w^2)$ is the \ac{ue} \ac{snr} at \ac{ofdm} symbol $n$, where $P_{\textnormal{UE}}(n)$ is the power that the \ac{ue} allocates at symbol $n$ and evenly distributes among the data subcarriers; in our simulations we set  $\textnormal{SNR}_{\textnormal{UE}} = 10 \, \textnormal{dB}$;
	\item $\textnormal{SNR}_{\textnormal{J}}(n) = P_{\textnormal{J}}(n) / (S \cdot \sigma_w^2)$ is the jammer \ac{snr} at \ac{ofdm} symbol $n$, where $P_{\textnormal{J}}(n)$ is the power that the jammer allocates at symbol $n$ and evenly distributes among the jammed subcarriers.
\end{itemize}
Moreover, for the Rayleigh case, we consider a block fading model and assume different channel realizations on different \acp{prb}. For \ac{urllc} type of traffic, we consider one packet per \ac{prb}, with $\rho = 0.48 \, \textnormal{bit/s/Hz}$.

\begin{figure}
	\centering
	\includegraphics[width=0.5\textwidth]{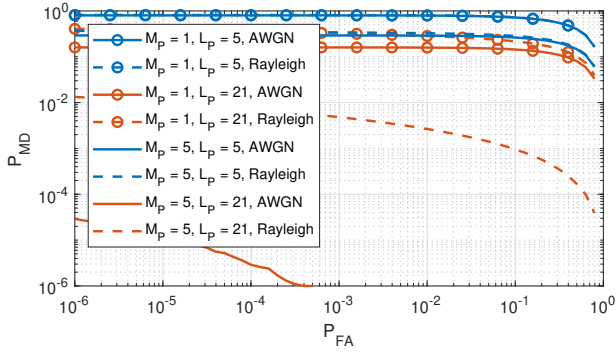}
	\caption{\ac{roc} curves for different $M_P$, $L_P$, and channel type. Here $\textnormal{SNR}_{\textnormal{J}} = 0 \, \textnormal{dB}$.}
	\label{fig:ROC_PJ_N14_DBconstant_PB5_DJconstantPRBs_PJ12-300_SNRJ0}
\end{figure}

Let's start with the performance evaluation of the defense strategy, in terms of \ac{md} probability as a function of the \ac{fa} probability, a.k.a. \ac{roc} curve. Fig. \ref{fig:ROC_PJ_N14_DBconstant_PB5_DJconstantPRBs_PJ12-300_SNRJ0} shows the \ac{roc} for $M_P=1,5$, $L_P=5,21$ (an almost narrow- and an almost wide-band jammer), $\textnormal{SNR}_{\textnormal{J}} = 0 \, \textnormal{dB}$, and for both \ac{awgn} and Rayleigh. First, we notice a huge performance improvement when using $M_P = 5$ when compared to $M_P = 1$, especially with $L_P = 21$ subcarriers and \ac{awgn}: in fact, while with $M_P = 1$ the \ac{md} is almost always above $10^{-1}$ for the considered range of target \ac{fa}, with $M_P = 5$ the detection performance strongly improves. Moreover, we also observe that while the narrow-band jammer ($L_P = 5$) is hardly detectable, the wide-band one can be easily spotted by the proposed method even if we have small jamming power. Finally, results show that detection in a \ac{awgn} scenario is far easier when compared to detection in a more random channel like the Rayleigh considered here.

\begin{figure}
	\centering
	\includegraphics[width=0.5\textwidth]{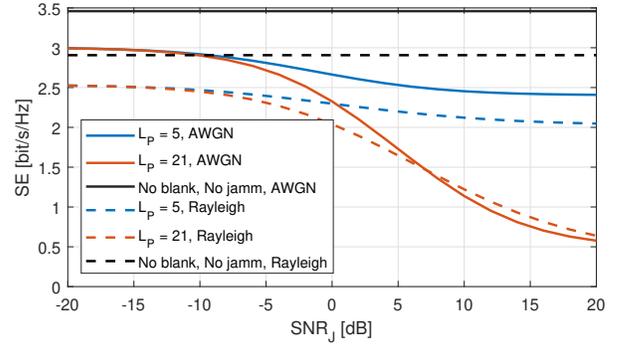}
	\caption{\Ac{se} versus $\textnormal{SNR}_{\textnormal{J}}$ for different $L_P$ and channel type. Here $M_P = 5$.}
	\label{fig:SE_PJ_N14_DBconstant_PB5_DJconstantPRBs_PJ12-300_SNRJ-20-20}
\end{figure}

To evaluate the performance degradation with a \ac{mbb} type of traffic, Fig. \ref{fig:SE_PJ_N14_DBconstant_PB5_DJconstantPRBs_PJ12-300_SNRJ-20-20} shows the \ac{se} as a function of $\textnormal{SNR}_{\textnormal{J}}$ with $M_P = 5$, for the almost narrow- and almost wide-band attack, and for a system with no blanking and no jamming that provides an upper bound to the proposed method. First, we notice, as expected, a small performance loss of the proposed method against the upper bound at very low jamming power because of the \ac{prb} blanking. Moreover, while the wide-band attack causes significant \ac{se} loss, especially for high $\textnormal{SNR}_{\textnormal{J}}$, the narrow-band attack, that resulted in Fig. \ref{fig:ROC_PJ_N14_DBconstant_PB5_DJconstantPRBs_PJ12-300_SNRJ0} to be more stealthy, only slightly limits the system \ac{se}. 

\begin{figure}
	\centering
	\includegraphics[width=0.5\textwidth]{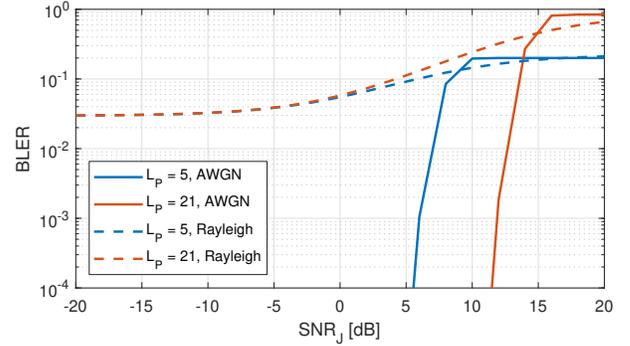}
	\caption{\ac{bler} versus $\textnormal{SNR}_{\textnormal{J}}$ for different $L_P$ and channel type. Here $M_P = 5$.}
	\label{fig:BLER_SNRJ_N14_DBconstant_PB5_DJconstantPRBs_PJ12-300_SNRJ-20-20_rate0.48}
\end{figure}

Concerning the \ac{urllc} type of traffic, Fig. \ref{fig:BLER_SNRJ_N14_DBconstant_PB5_DJconstantPRBs_PJ12-300_SNRJ-20-20_rate0.48} shows the \ac{bler} as a function of $\textnormal{SNR}_{\textnormal{J}}$ for $L_P=5,21$ jammed \acp{prb} and $M_P=5$ blanked \acp{prb}. In the \ac{awgn} channel, we observe that with limited jamming power, a narrow-band attack allows the jammer to strongly degrade the performance and at the same time avoid the blanked \acp{prb}. But, as its power increases, the \ac{bler} reaches a saturation value, which depends on the probability of intersection between blanked and jammed \acp{prb}, and for higher $\textnormal{SNR}_{\textnormal{J}}$ it should switch to a wide-band attack. When looking at the Rayleigh case, we observe that, in general, the system performs worse than the \ac{awgn} case
.

\begin{figure}
	\centering
	\includegraphics[width=0.5\textwidth]{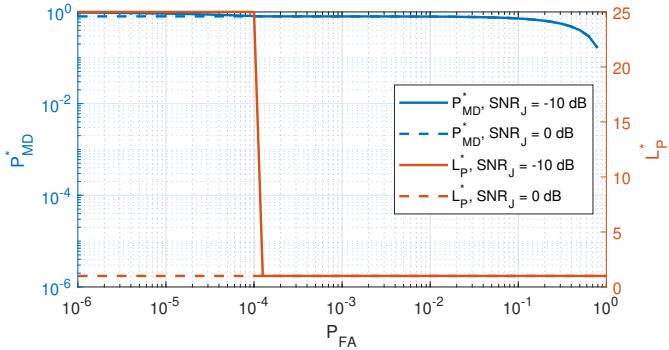}
	\caption{Optimal $P_{\textnormal{MD}}$ (left y-axis) and optimal $L_P$ (right y-axis) versus $P_{\textnormal{FA}}$ for different $\textnormal{SNR}_{\textnormal{J}}$. Here $M_P = 5$ and \ac{awgn} is considered.}
	\label{fig:ROC_PJ_N14_DBconstant_PB5_DJconstantPRBs_PJ12-300}
\end{figure}

In Fig. \ref{fig:ROC_PJ_N14_DBconstant_PB5_DJconstantPRBs_PJ12-300} we evaluate the jammer strategy for \ac{md} maximization proposed in Section \ref{sec:jamming_strategies_A} by showing for $\textnormal{SNR}_{\textnormal{J}} = -10, 0 \, \textnormal{dB}$ the best \ac{md} probability achievable by the attacker and the corresponding number of jammed \acp{prb} to achieve it. For the low power jammer, i.e., $\textnormal{SNR}_{\textnormal{J}} = -10 \, \textnormal{dB}$, we notice that if the defender's target \ac{fa} probability is $P_\textnormal{FA} \gtrsim 10^{-4}$, the best approach for the jammer is to perform a narrow-band attack; this happens because the defense tends to accept the $\mathcal{H}_1$ hypothesis more easily, and therefore the attacker tries to avoid the blanked \acp{prb} by transmitting on a smaller number of \acp{prb}. On the other hand, for $P_\textnormal{FA} \lesssim 10^{-4}$, the jammer best strategy is a wide-band attack because, in this way, it evenly distributes its power among all the subcarriers
. On the contrary, with a higher jamming power, i.e., $\textnormal{SNR}_{\textnormal{J}} = 0 \, \textnormal{dB}$, we see that the optimal strategy is the narrow-band attack for the entire \ac{fa} probability interval that we consider.

\begin{figure}
	\centering
	\includegraphics[width=0.5\textwidth]{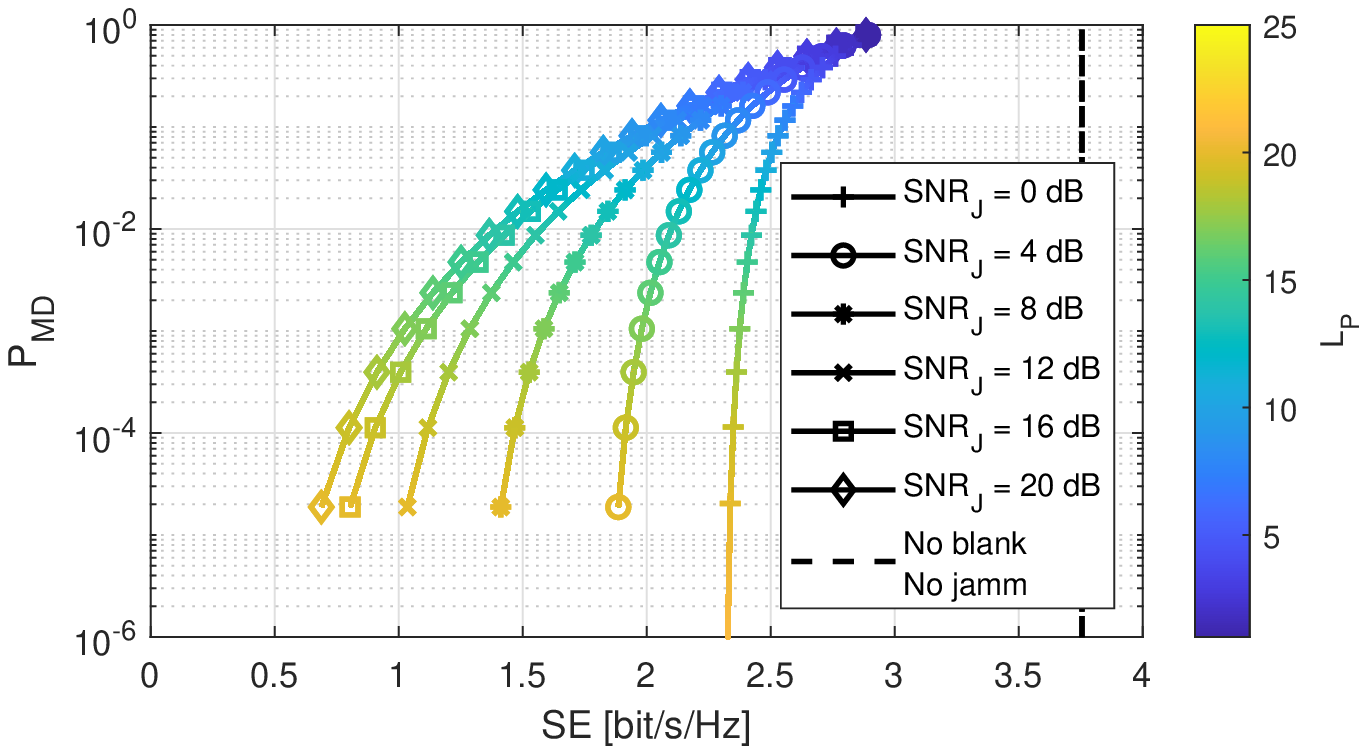}
	\caption{$P_{\textnormal{MD}}$ versus \ac{se} for different $\textnormal{SNR}_{\textnormal{J}}$. Here $P_{\textnormal{FA}} = 10^{-3}$, $M_P = 5$, and \ac{awgn} is considered.}
	\label{fig:PMD_SE_SNRJ_N14_DBconstant_PB5_DJconstantPRBs_PJ12-300_SNRJ0-20}
\end{figure}

In Section \ref{sec:jamming_strategies_B} we showed that the best strategy to minimize the \ac{se} is a wide-band attack, while in Fig. \ref{fig:ROC_PJ_N14_DBconstant_PB5_DJconstantPRBs_PJ12-300} we learned that, on the contrary, in many cases the narrow-band attack is the best strategy to maximize the \ac{md}, thus suggesting a trade-off between \ac{md} probability and \ac{se}. In Fig. \ref{fig:PMD_SE_SNRJ_N14_DBconstant_PB5_DJconstantPRBs_PJ12-300_SNRJ0-20} we evaluate this trade-off by showing the \ac{md} probability versus the \ac{se}, for different values of $L_P$ and $\textnormal{SNR}_{\textnormal{J}}$, and for $P_{\textnormal{FA}} = 10^{-3}$. The optimal situation for the attacker would be to achieve high $P_{\textnormal{MD}}$ and low \ac{se}, but, for the considered range of $\textnormal{SNR}_{\textnormal{J}}$, there is a maximum that can be achieved and, depending on its objective, the jammer needs to give up on \ac{se} reduction if it wants to increase the \ac{md} and viceversa.

\begin{figure}
	\centering
	\includegraphics[width=0.5\textwidth]{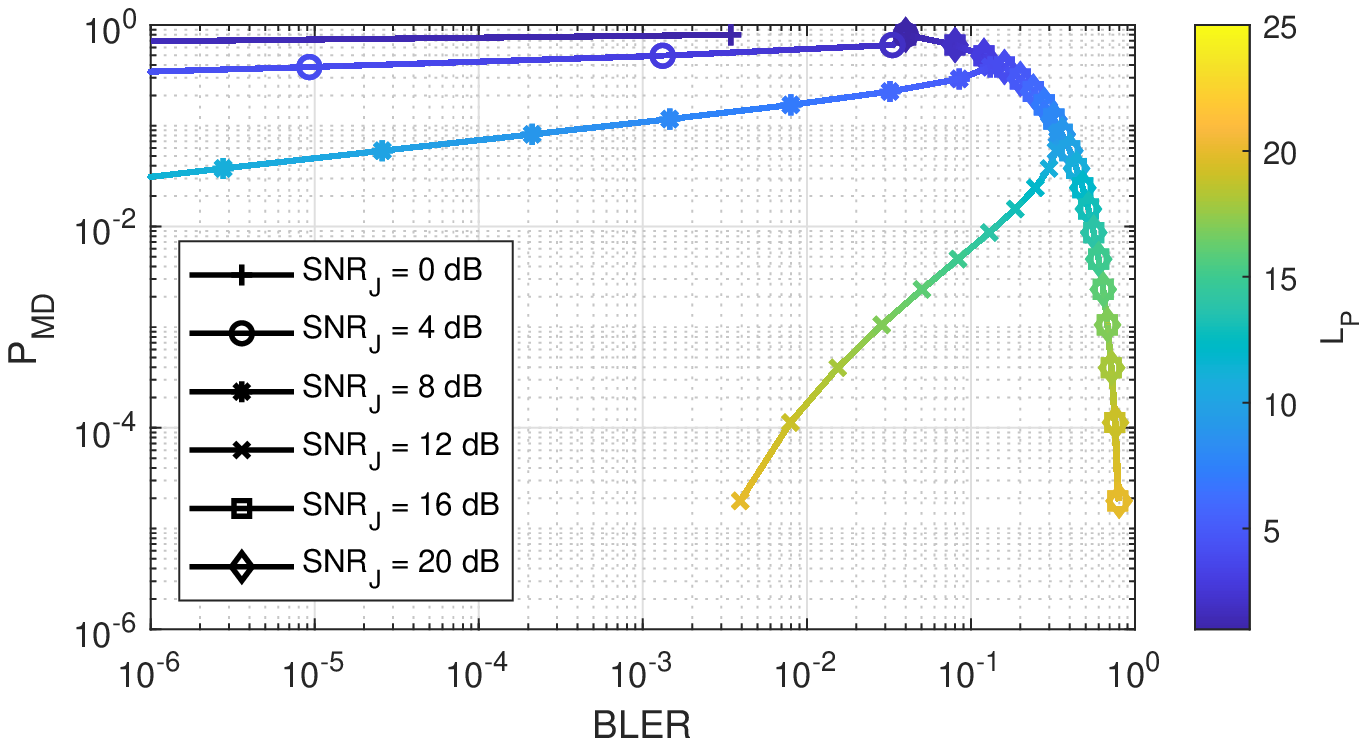}
	\caption{$P_{\textnormal{MD}}$ versus \ac{bler} for different $\textnormal{SNR}_{\textnormal{J}}$. Here $P_{\textnormal{FA}} = 10^{-3}$, $M_P = 5$, and \ac{awgn} is considered.}
	\label{fig:PMD_BLER_SNRJ_N14_DBconstant_PB5_DJconstantPRBs_PJ12-300_SNRJ0-20}
\end{figure}

Finally, Fig.  \ref{fig:PMD_BLER_SNRJ_N14_DBconstant_PB5_DJconstantPRBs_PJ12-300_SNRJ0-20} considers the \ac{bler} maximization problem of Section \ref{sec:jamming_strategies_C} and shows the \ac{md} probability versus the \ac{bler}, for different values of $L_P$ and $\textnormal{SNR}_{\textnormal{J}}$, and for $P_{\textnormal{FA}} = 10^{-3}$. These results show that if the jammer has low power, for instance with $\textnormal{SNR}_{\textnormal{J}} = 0 \, \textnormal{dB}$, it cannot achieve high \ac{bler} and at the same time stay undetected. On the other hand, by looking at the top-right region of the plot, we observe that a jammer with a sufficiently high power can use a narrow-band attack to achieve high \ac{bler} and high $P_{\textnormal{MD}}$
.

\section{Conclusions}\label{sec:conclusions}

In this paper, we considered the problem of jamming detection for \ac{5g}-and-beyond in Industry 4.0 scenarios 
and 
%
designed a method based on pseudo-random blanking of subcarriers in an \ac{ofdm} system
. We then considered a smart jammer following three types of strategies: remain as stealthy as possible, maximize the damage to \ac{mbb} communication, and maximize the disruption of \ac{urllc} type of traffic. Results show that, while for a \ac{mbb} traffic the jammer has to compromise between \ac{md} and \ac{se}, with \ac{urllc} traffic, a smart jammer with sufficiently high power can achieve good results in reaching both high values of \ac{md} probability and \ac{bler}.
Future works include 
performance evaluations in a \ac{3gpp} compliant Industry 4.0 scenario.

\balance 

\bibliographystyle{IEEEtran}
\bibliography{IEEEabrv,bibliography}

\end{document}